\documentstyle[12pt,aaspp]{article}
\def\lsim{\raise2.90pt\hbox{$\scriptstyle
<$} \hspace{-6pt}\lower.5pt\hbox{$\scriptscriptstyle\sim$}\; }
\def\bi#1{\hbox{\boldmath{$#1$}}}
\def\ltsima{$\; \buildrel < \over \sim \;$}
\def\lsim{\lower.5ex\hbox{\ltsima}}
\def\gtsima{$\; \buildrel > \over \sim \;$}
\def\gsim{\lower.5ex\hbox{\gtsima}}

\begin{document}

\title{Cosmography and  
Power Spectrum Estimation: a Unified Approach}
\author{Uro\v s Seljak}
\affil{ Harvard Smithsonian Center For Astrophysics, Cambridge, MA 02138 USA
}

\begin{abstract}
We present a unified approach to the problems of 
reconstruction of large-scale structure distribution in the universe and 
determination of the underlying power spectrum. These have
often been treated as two separate problems
and different analysis techniques have been developed for both.
We show that there exists
a simple relation between the optimal solutions to the two problems, 
allowing to solve for both
within the same formalism.
This allows one to apply computational techniques developed for
one method to the other, which often leads to a significant reduction 
in the computational time. 
It also provides a self consistent treatment of linear reconstruction by 
first computing the power spectrum from the data itself.
\end{abstract}

\newpage

\section{Introduction}

The issue of optimal power spectrum reconstruction from a 
set of noisy and sparsely sampled data has 
recently received a lot of attention in the field of 
large scale structure (LSS) and cosmic microwave background (CMB)
anisotropies (\cite{kbj,hama,teg,tegb,vs}). 
Most of this work has been inspired by the existing and 
forthcoming large data sets (COBE, MAP, Planck, SDSS, 2DF etc.).
The quality and amount of information in these surveys is or 
will be so large 
to allow one to make high precision tests of cosmology. 
The precision to which one can test 
cosmological models is limited to the accuracy with which one
can extract useful information from the data and so it is clearly
important to develop statistical techniques that will allow one to minimize 
the loss of information from the data. This can be formalized 
with the use of information theory, which allows one to define 
the requirements for an optimal estimator (\cite{tth}).
The central object in this discussion is the Fisher information 
matrix, defined as the ensemble average of the 
matrix of second derivatives of the (minus) 
log-likelihood function with respect to the parameters we wish to estimate.
Its inverse provides a minimum bound for 
covariance matrix of the parameters, known in the literature as the 
Cram\' er-Rao inequality (e.g. Kendall and Stuart 1969). 
The natural question to ask is which 
methods allow us to reach this theoretical limit.
Maximum-likelihood is obviously a preferred method in this sense, as
it allows one to directly explore the likelihood function, finding its
maximum and second derivatives around it.
Unfortunately, to compute the likelihood function one has to invert the
data matrix for each set of parameters, which becomes computationally 
prohibitive for large data sets. While linear compression methods,
such as Karhunen-Lo\` eve (\cite{bond,bs,tth,vs}), 
allow to reduce the size of the system
if the data are oversampled, the compression is not sufficient for future 
surveys with millions of data points. The problem is present in
all linear methods, because
they keep the information on individual measurements, while what 
one wants to extract from a stochastic gaussian process (thought 
to be a good model for matter distribution in 
the universe on large scales)
is a quadratic quantity in the data.
This can under the assumption of isotropy
be averaged over all the pairs contributing to the same angular 
separation or mode amplitude, which then contains all the information
about the cosmological model.
One should therefore be able to compress the data to a correlation 
function or power spectrum
estimator without losing any information in the case of
a gaussian random processes. 
Recently, a quadratic estimator for the power spectrum has been proposed 
that allows one to reach the limit
given by Fisher matrix directly (\cite{hama}). 
The explicit form of the estimator is particularly simple in its pixelized
form if one assumes gaussian distribution of the data (\cite{teg}) and
it has been shown that it gives equivalent results as the likelihood 
analysis (\cite{kbj}).
While the gaussian assumption breaks down because of nonlinear evolution
on small scales, it should
be a good approximation for the CMB data and
for galaxy and weak lensing surveys on large scales.
Even in the nonlinear regime
the gaussian estimator remains a useful approximation, being much 
simpler to compute than the general expression. The main complication 
in the nonlinear regime is
that the covariance matrix depends on the 4th moment, which 
has to be computed either using Monte 
Carlo N-body simulations or perturbation theory.
Although both 
correlation function and power spectrum have been used in the past
to measure the statistical correlations between the data, there are
several advantages to the power spectrum analysis. For example,
power spectrum can exhibit
physical processes, such as acoustic oscillations in CMB, more clearly
than the correlation function. More importantly,
the observations of non-scalar quantities, such as velocity flows, 
weak lensing or polarization of CMB,
can be more easily related to the underlying 
theory in signal eigenmode space than in real space, where the correlation
functions are correlated (\cite{kaiser,seljak97a}).

Computing the power spectrum from a survey is not the only 
interesting question that one can address with survey data.
Often one wants to reconstruct 
the true underlying distribution, given the noisy and sparsely sampled set 
of observations. Such a reconstructed image can be used to search for  
individual objects, such as cold and hot spots in CMB maps or
clusters, superclusters, filaments and
walls, 
to count the number density 
of such objects as a function of their mass or luminosity 
or to look at their individual morphologies and substructure.
Using a reconstructed map one can study the topology of the universe or
compare different observations of the same region. 
A simple and powerful method of reconstruction is
Wiener filtering (WF), 
which minimizes the variance in the class of linear estimators 
(\cite{rp}). Just like the minimum variance
estimator WF does not require the data to be gaussian distributed, 
but can be shown to be optimal in this limit, since it coincides with
the maximum posterior probability estimator. As such it is clearly 
the method of choice in the CMB and LSS reconstruction, where the
deviations from gaussianity are either 
small or negligible and has successfully been applied in these cases
(\cite{bunn,tegetal,zar} and references therein). 
The main shortcoming of WF is that it requires
as input the power spectrum of the underlying field and this is often 
not known in advance. In the absence of any external information it 
has to be obtained from the data itself. 
We will show in this paper how to optimally
extract the power spectrum from the WF map, thus providing 
a self-consistent treatment of WF. 

The two analysis techniques, power spectrum estimation and image 
reconstruction, have so far been disconnected in the literature. 
Naive power spectrum estimation from the WF maps leads to biased
results (a well known consequence of WF to return zero in the regimes
of low signal to noise) and was not used for this reason. Optimal 
power spectrum estimation techniques are generally thought to 
require very specific analysis techniques, too different from 
image reconstruction to be able to address both problems 
simultaneously. It would however be useful if a single analysis 
method produced both power spectrum estimation and image 
reconstruction. It would be even more useful if such analysis 
would not increase the computational time for either of the methods.
This would then be the preferred method even if 
only one application was the primary objective of the analysis. 
In this paper we show that there is a simple connection between the
two methods, which gives the two desired properties above. This 
provides a unified approach to the two analysis techniques 
and allows one to obtain both image reconstruction and minimum variance
power spectrum estimator within the same analysis. Moreover, this connection
allows one to translate various calculational methods from one 
application to the other. 
For example, transformation to Fourier space is 
commonly employed 
in WF analysis and leads to a significant
reduction of computational time for some problems. 
The same transformation can also be used for minimum variance
power spectrum estimator. 
Conversely, approximation techniques developed for
minimum variance
power spectrum estimator can also be applied to WF analysis and reduce 
its computational time significantly, allowing large maps to be 
reconstructed. 

In this paper we concentrate on deriving a set of tools  
for image reconstruction
and power spectrum estimation, paying special attention to the 
connections between the two. The main goal is to present a self-contained
treatment of the two problems, so some of the ideas presented will not be 
fundamentally new. Given the many possible applications of these methods
to the existing and forthcoming LSS and CMB surveys
it seems important to present them with a unified treatment.
Among the possible applications 
are galaxy and weak lensing surveys, 
peculiar velocity flows,  
Lyman $\alpha$ forest from quasar spectra 
and CMB temperature and polarization maps. All these have a common
feature of mapping the universe on large scales, where the techniques
explored here work best.
Some of these applications will be presented elsewhere. 
The outline of this paper is the following:
in section 2 we review the minimum variance 
estimator. In section 3 we 
first review the WF reconstruction 
procedure and then show its connection to the 
minimum variance estimator. Section 4 discusses the reconstruction 
and estimation in signal eigenmode basis, where the calculations 
often become less computationally intensive. 
Section 5 applies the general expressions
from previous sections to the case of plane wave expansion, which is of 
particular importance to the LSS surveys and where translational invariance
and Fast Fourier transform (FFT) techniques can be utilized to reduce 
the computational time. 
This is followed by the general discussion
in section 6. Treatment of external 
parameters and other constraints is left for appendix.

\section{Minimum Variance Power Spectrum Estimator}
In this section we review the formalism related to the minimum 
variance estimator.
Although this section has no major new 
results in itself, it is included here for completeness, as it
establishes the notation used in the rest of the paper.
For further details see \cite{kbj,hama} and \cite{teg}. 

Let us assume that we measure the quantities $d(\bi{r}_i)$
at $N$ positions $\bi{r}_i$, either in the sky or in space.
We arrange these into a vector 
$\bi{d}=\{d(\bi{r}_i)\}(i=1,...,N)$. 
The measured quantities $\bi{d}$ can be either
scalars, such as temperature anisotropy or density 
perturbation, components of a vector (e.g. radial peculiar velocity) 
or components of a
tensor, such as polarization of microwave background or
galaxy ellipticity in the weak lensing analysis. 
Each measurement consists of a signal and noise 
contributions, $\bi{d}=\bi{Rs}+\bi{R_bs_b}+\bi{n}$. 
Here $\bi{R}$ is 
a $M \times N$ response matrix and $\bi{s}=\{s_\lambda\}(\lambda=1,...,M)$ 
are the underlying field coefficients that we wish to estimate. True
underlying field has always infinite number of coefficients, but for 
computational reasons only a finite number of these can be estimated. 
For this reason
we group all the modes that we do not estimate 
into $\bi{s_b}$, with $\bi{R_b}$ the corresponding response matrix.
The underlying field is parametrized with covariance matrices 
$\bi{S}=\langle \bi{ss}^{\dag}\rangle$ and 
$\bi{S_b}=\langle \bi{s_bs_b}^{\dag}\rangle$. The noise vector
$\bi{n}=\{n_i\}(i=1,...,N)$ 
is parametrized with the 
noise covariance matrix $\bi{N}=\langle \bi{nn}^{\dag}\rangle$.
This noise matrix is often
known a priori, uncorrelated with the signal
and diagonal in real space (the former is not exactly true if the 
noise is 
coupled to the signal, as is the case for shot noise in galaxy clustering; 
the latter is not true if $1/f$ noise
is important, as is the case in some, but not all, CMB experiments).
When we take the 
response functions to be the mode expansion coefficients then the
signal covariance matrices $\bi{S}$ and $\bi{S_b}$ are
also diagonal and only depend on the 
amplitude of the mode. The correlation matrix of the data is given by
\begin{equation}
\langle \bi{dd}^{\dag} \rangle
\equiv \bi{C}=\bi{RSR}^{\dag}+\bi{C_b}+
\bi{N}=\sum_l\Theta_l\bi{Q}_l+\bi{N}. 
\label{corr}
\end{equation}
Here $\bi{RSR}^{\dag}=\sum_{l=1}^{l_{max}}\Theta_l \bi{Q}_l$
was simplified by summing over all the $M_l$ 
modes $\bi{s_l}=\{s_{m_l}\}(m_l=1,...,M_l)$, whose mode amplitude
contributes to the portion of the power 
spectrum parametrized with $\Theta_l$, so that 
$\langle s_{m_l}s_{m_l}^* \rangle=\Theta_l$. 
The aliasing term $\bi{C_b}$ was similarly 
rewritten from individual modes to
a sum (or integral) over the power spectrum,
$ \bi{C_b}=
\bi{R_bS_bR_b}^{\dag}=\sum_{l=l_{max}+1}^{\infty}\Theta_l \bi{Q}_l$.
We can introduce a 
projection matrix $\bi{\Pi}_l$, which consists of ones along the diagonal 
corresponding to the $M_l$ modes and zeros otherwise, in terms of which
$\bi{Q}_l=\bi{\Pi}_l\bi{R}\bi{R}^{\dag}\bi{\Pi}_l$.
For example, in the spherical harmonic
decomposition of temperature anisotropies in the sky we can decompose 
them as $\Delta(\bi{n_i})=\sum a_{lm} Y_{lm}(\bi{n}_i)$, 
where $Y_{lm}(\bi{n}_i)$ are the spherical harmonics
in the direction $\bi{n}_i$. If we choose to model the field with all
the modes up to $l_{max}$ then the underlying field is 
$\bi{s}=\{s_\lambda\}=\{a_{lm}\}(\lambda=l^2+l+m+1, l=1,...,l_{max},
m=-l,...,l)$ and $\bi{s}_b=\{a_{lm}\}(l=l_{max}+1,...,\infty)$, while
the response matrices are 
$R_{\lambda,i}={Y_{lm}(\bi{n}_i)}(l=1,...,l_{max})$ and 
$R_{b(\lambda,i)}={Y_{lm}(\bi{n}_i)}(l=l_{max}+1,...,\infty)$. 
The correlation
matrix has contributions from all $l$,
$\langle a_{lm}a_{l'm'}^* \rangle= C_l \delta_{ll'}\delta_{mm'}$
and we can identify $\Theta_l=C_l$. Projection matrix 
$\bi{\Pi}_{l,\lambda,\lambda'}$ has ones for $\lambda=\lambda'=
l^2+l+m+1, m=-l,...,l$
and zeros otherwise. Finally
$Q_{l,ij}=\sum_m Y_{lm}(\bi{n}_i)Y^*_{lm}(\bi{n}_j)$ and using the 
addition theorem for spherical harmonics we find
$Q_{l,ij}={2l+1\over 4\pi} P_l(|\bi{n}_i-\bi{n}_j|)$, 
where $P_l$ is the Legendre polynomial. 
The above notation is however more general and allows to parametrize 
the power spectrum with fewer parameters. 
If the survey size $D$ is such that $\Delta l \sim 1/D \gg 1$
or if we have reasons to believe that 
the underlying power spectrum is smooth then we 
may group together neighbouring mode amplitudes in the power spectrum
and parametrize them with a 
single parameter $\Theta_l$, where now $l$ 
simply counts the parameters we are 
trying to estimate and $l_{max}$ is their total number. 
The corresponding $Q_l$ is still a sum over squares of
all individual modes 
that contribute to $l$-th parameter $\Theta_l$.
For surveys with different geometry a 
different expansion (e.g. in plane waves) may be more appropriate,
but it can nevertheless be written in the above form (replacing
some of the sums with integrals).
Of course, other parametrizations of the power spectrum are also 
possible, for example its amplitude and slope, 
but for our purposes parametrization in terms of its value at a
given mode amplitude will be the most useful form, 
since it can be directly related to the underlying field $\bi{s}$
that we wish to estimate. Other parametrizations with fewer
parameters could however be
used for the power spectrum $\bi{S}_b$ that characterizes the modes
$\bi{s}_b$ we do not estimate, especially 
when we do not have sufficient information on it, but we nevertheless
wish to estimate its influence on the parameters we are estimating.

Given a set of measurements $\bi{d}$ we wish to find the most probable
set of parameters $\bi{\Theta}$. For gaussian theories all the information
from the data is contained in the likelihood function
\begin{equation}
L(\bi{d} \vert \bi{\Theta})=(2\pi)^{-N/2} \det(\bi{C})^{-1/2} \exp(-{1 \over 2}
\bi{d}^{\dag} \bi{C}^{-1}\bi{d}),
\label{lik}
\end{equation}
where $\bi{C}$ implicitly depends on the parameters $\bi{\Theta}$.
To find the most probable set of parameters one needs to find the
maximum of the likelihood function above \footnote{All the discussion 
here is also valid from a Bayesian point of view, provided that the 
likelihood is interpreted as a posteriori probability. In the case of
uniform prior in the parameters 
the above expressions remain the same, otherwise the
appropriate prior has to be added (e.g. \cite{kbj}).}.
By Taylor expanding the
log of likelihood function 
$L(\bi{d} \vert \bi{\Theta})$ 
around the model parameters $\hat{\bi{\Theta}}$ that maximize it we
obtain the matrix of second derivatives, called the curvature matrix.
Because the likelihood function is nonlinear as a function of parameters
one has to sample it in a number of parameters
to find the location of the maximum and the second derivatives around it. 
Rather than 
computing the curvature matrix, which is computationally intensive,
one often computes its ensemble average, called the 
Fisher matrix 
\begin{equation}
F_{ll'}=-\left\langle {\partial^2 \ln L(\bi{\Theta}; \bi{d}) 
\over \partial \Theta_l \partial \Theta_{l'}}
\right\rangle_{\hat{\bi{\Theta}}}.
\end{equation}
Brackets denote ensemble averaging. When the likelihood
function can be approximated as a gaussian around the maximum and 
the maximum likelihood estimator is close to the true value then
one can interpret the inverse of the Fisher matrix $\bi{F}^{-1}$ as an
estimate of the covariance matrix of the parameters $\hat{\bi{\Theta}} $,
\begin{equation}
\langle \hat{\bi{\Theta}}\hat{\bi{\Theta}}^{\dag}\rangle -
\langle \hat{\bi{\Theta}}\rangle\langle \hat{\bi{\Theta}}\rangle^{\dag}
=\bi{F}^{-1}.
\label{minvar}
\end{equation}
Note that the gaussian assumption of the likelihood function does not 
rely on the gaussianity of the data: likelihood function will be 
gaussian around the maximum provided that sufficient independent 
modes contribute to each $\Theta_l$, by central limit theorem.
Cram\' er-Rao inequality (e.g. Kendall and Stuart 1969) 
states that the best an unbiased estimator
can do is to reach the limit given by the Fisher matrix, i.e. 
$\sigma(\hat{\Theta_l}) \geq (\bi{F}^{-1})^{1/2}_{ll}$.
Clearly, brute force maximum likelihood search will achieve this limit, but 
is not the fastest method and computational cost in many cases
becomes prohibitively expensive. 

The computational difficulties mentioned above lead 
a number of authors to investigate a quadratic 
estimator that allows to find the maximum of the likelihood
more rapidly (\cite{hama,kbj,teg}). 
For the case where the data are gaussian distributed with zero mean 
the estimator is
\begin{equation}
\hat{\Theta}_l={1 \over 2}\sum_{l'}F^{-1}_{ll'}[\bi{d}^{\dag} \bi{C}^{-1}\bi{Q}_{l'}
\bi{C}^{-1} \bi{d}-b_{l'}].
\label{estimator}
\end{equation}
Here $b_l$ is the noise term, which can be obtained by computing 
ensemble average of $\bi{d}^{\dag}\bi{C}^{-1}\bi{Q}_l
\bi{C}^{-1} \bi{d}$, assuming $\Theta_l=0$ for $l\le l_{max}$. This
gives
\begin{equation} 
b_l=tr[\bi{(N+\sum_{l_{max}+1}^\infty \Theta_l 
\bi{Q}_l) C}^{-1}\bi{Q}_l\bi{C}^{-1}].
\label{bl}
\end{equation}
The term $\sum_{l_{max}+1}^\infty \Theta_l 
\bi{Q}_l$ has been left out in previous work, because it was implicitly
assumed that all the relevant modes are being estimated. In practice this
is not the case when the data are sparsely sampled and have low noise. 
In this case the mixing between the modes 
aliases power 
from small scales to large scales and this term may dominate over the usual
noise term. One could argue of course that if the data still have some 
excess of power above the noise then one should try to estimate this 
power rather than treat it as an additional noise term. There are 
several instances when this may not be practical. For exaqmple, when the
sampling is very sparse (as for example in supernova measurements of
velocity field), then there are simply too many small scale
modes to include them all in the parametrization. Another possibility is 
that we have already measured the power on small scales accurately 
by some other means,
so we do not need to include those modes for estimation. In both 
cases one can 
include the effect of small scale modes as an additional noise term, which 
has to be included in equation (\ref{bl}) for the estimator to be unbiased.
Even when not in such clear situations
one should always check whether the aliasing is 
important by computing its contribution, assuming a reasonable (or already
measured) power spectrum for the unestimated small scale modes.
Fisher matrix $\bi{F}$ is given by 
\begin{equation}
F_{ll'}={1 \over 2}tr(\bi{Q}_{l}\bi{C}^{-1}\bi{Q}_{l'}\bi{C}^{-1}).
\label{fisher}
\end{equation}
This estimator 
has the nice properties of being unbiased, $\langle \hat{\Theta} \rangle=
\Theta$,
and minimum variance  as defined in equation (\ref{minvar}) (\cite{teg}).

The estimator remains unbiased when the gaussian assumption is dropped 
(since its mean depends only on the second moment of the distribution) and
so one can continue to use it in the nonlinear regime. In this case 
the estimator as written above 
is no longer minimum variance, although its 
computational simplicity compared to the general case (\cite{hama})
outweights the loss of information 
in practical applications. The more important difficulty in this case is
that the covariance matrix
for the estimators is no longer given by the inverse of $\bi{F}$ as 
defined in equation (\ref{fisher}), because it depends on the 
four point function of the distribution, 
which does not vanish for the nongaussian 
process. It has to be computed explicitly, for example by 
using Monte Carlo methods on N-body simulations or by perturbation theory. 
Provided this is properly computed then the estimator above 
remains a useful working model for the power spectrum estimation 
even in the nongaussian regime. 

Although we wrote the 
optimal estimator in equation (\ref{estimator}) in terms of $\bi{F}^{-1}$, 
inverting the Fisher matrix may not be stable, when the matrix is 
nearly singular. This happens if the binning of the power 
spectrum is too narrow for the survey geometry, i.e. $\Delta l \ll 
1/D$, where $D$ is the size of the survey. Two simple solutions 
around this problem are:
\newline a) parametrize the power spectrum with 
fewer parameters, essentially coarse graining the spectrum. This
way the Fisher matrix is not singular and can be inverted. One
has to be careful to include the effects of a changing power 
spectrum across the band. 
Each mode of $\bi{R}$ contributing to $\bi{Q}_l=\bi{\Pi}_l\bi{R}\bi{R}^{\dag}\bi{\Pi}_l$
gives
an estimate of $\bi{S}$ at that mode amplitude and if this is changing 
across the band then it is better to estimate a quantity that is flat
across the band. To do this we divide each mode component in
$\bi{d}^{\dag} \bi{C}^{-1}\bi{Q}_l
\bi{C}^{-1} \bi{d}$ with the prior power spectrum (one that we believe
describes well the change of power spectrum across the band), 
evaluated at the amplitude of that mode.  
In the end we put this term back by multiplying the power spectrum 
estimate with the prior power spectrum evaluated
at the mode amplitude to which we want to assign the estimator.
This procedure will guarantee an unbiased result (see \cite{kbj} for
a similar treatment) and can be quite important when 
the power spectrum is steep and binning coarse.
\newline 
b) do not attempt to invert $\bi{F}$, but instead quote filtered
estimators $N_l\bi{F}\hat{\bi{\Theta}}$, 
where $N_l=(\sum_{l'}F_{ll'})^{-1}$.
The filtering function 
$N_l\bi{F}$ is a bell shaped function in $l'$ for a given $l$
and its width around $l$ tells us the spectral resolution at that $l$.
Covariance matrix for this estimator is given by $N_lN_{l'}F_{ll'}$
(\cite{teg,kbj}). 
More complicated
methods which make the power spectrum estimators uncorrelated 
have also been proposed (\cite{hamb,teg,kbj}). They still require 
one to report the full filtering matrix and are useful mainly 
if one wishes to further compress the estimators, for example
for the purpose of plotting them on a graph. 

Once a power spectrum estimate has been obtained
one wants to test different cosmogonies or extract cosmological 
parameters from it. 
The interesting question is whether the optimal 
way to perform this step is using the data or the quadratic estimators.
A comparison between the two methods should help to elucidate the 
similarities and differences between the two. In the 
linear case we use the actual data in the analysis and try to maximize 
the likelihood function. This should be done using the full probability
distribution of the data, but in practice for large data sets 
one can only perform the analysis 
under the assumption that the data are gaussian distributed. One
then computes 
the likelihood function using the
correlation matrix of the model one is testing and compares it to the 
maximum likelihood value. In the qudratic case one first compresses the 
data to the power spectrum estimators, which are quadratic in the data,
and performs the likelihood analysis using the covariance matrix of these
estimators. Here again one should use covariance matrix obtained from
the model
one is testing and, for consistency, do the power spectrum analysis 
using that model as well.
While the actual values of power 
spectrum estimators do not depend sensitively on the assumed model,
covariance matrix does if one is limited by the
sampling variance, because in this limit the error on the estimate is
dominated not by the noise but by the assumed power spectrum.
It should be emphasized
that in both cases to test a model one has to use the covariance matrix
from that model, not from the best fitted model. 
In general one therefore 
cannot attach model independent error bars to the power spectrum 
estimator for the purpose of model testing. 
If we assume that the covariance matrix has been calculated using the 
model one is testing then the only difference between 
the linear and quadratic approach lies in the gaussian assumption
for the model entering 
the likelihood function, either the actual data
or quadratic 
power spectrum estimators. The answer then depends on the particular 
application: if we have reasons to believe the data are gaussian 
distributed (as should be the case e.g. for CMB anisotropies), then linear analysis 
is clearly correct. For quadratic analysis the distribution for a given 
power spectrum estimator $\Theta_l$ is roughly 
$\chi^2(M_l)$, where $M_l$ is the effective number of independent 
modes contributing
to $\Theta_l$. Only if $M_l$ is large can this distribution 
be approximated as
a gaussian by central limit theorem. On the other hand, if the distribution
of individual data points is not gaussian the linear analysis 
is not correct (or becomes much more complicated if nongaussian likelihood
function is used), 
while the distribution of a quadratic estimator can still be gaussian 
in the same large $M_l$ limit. Of course in this case the covariance 
matrix for the estimators will not be given by equation (\ref{fisher}), 
which was derived under the gaussian assumption for the data, 
but provided that 
this matrix can be obtained by some other means the gaussian distribution
for the quadratic estimators is correct. 
The answer therefore depends on the nature
of the data and on the size of the survey. In general, for estimation on
large scales which are sampling variance limited and likely to 
be gaussian quadratic 
estimator will be the least reliable, but since these modes also have the 
largest errors attached the overall error may not be very 
important, at least
for finding the most probable model. The effects of the gaussian assumption 
are somewhat more important for setting the confidence limits, where the 
tails of distribution are important (\cite{kbj}).

The main advantage of the quadratic estimator is that it 
compresses the data to a small number of values and their covariance 
matrix, which are then easier to manipulate than the original data set.
The compression here can be truly enormous, from millions of measurements
to hundreds of power spectrum estimates in the case of future surveys.
As we argued above for the purpose of parameter extraction 
power spectrum covariance matrix cannot be assigned
in a model independent way
and in principle one would have to repeat the quadratic estimator
for each model one is testing. 
For practical purposes this is not necessary and one can 
approximately attach meaningful error bars to the power spectrum 
estimators. Since
we assumed that the power spectrum is known,
whereas in reality this is something we wish to estimate, one is tempted
to iterate the power spectrum estimation 
until the convergence is reached.  
This however will not be the best
solution, because if the estimators have large sampling variance 
then some will be low and some high and those that are low will 
also have a smaller attached sampling error, since it was based on 
the assumption that the true power spectrum is the measured 
one. This will therefore bias any subsequent estimation
of parameters. Two ways to resolve this problem are either to 
group the parameters so that the sampling variance is reduced
or to find a smooth curve that goes through the power spectrum estimators
and use that as the assumed power spectrum in the next step of 
iteration. 
The criterion for the smoothness of the curve should be based
on the requirement that $\chi^2=(\bi{\Theta}-\hat{\bi{\Theta}})^{\dag}
\bi{F}^{-1}
(\bi{\Theta}-\hat{\bi{\Theta}}) \approx l_{\rm max}$, where $\bi{\Theta}$
is the parametrization of the smooth power spectrum. In other words, 
the assumed power spectrum
should have sufficient detail to be reasonably close to the 
measured values, yet it should not attempt to match those values more
accurately than permitted by the attached errors. Since the covariance
matrix in $\chi^2$ depends on the assumed power spectrum in previous step
it may be necessary to repeat the iteration two more times, to a 
total of three estimations.

\section{Wiener Filter Reconstruction}
The goal of image reconstruction deals with the following problem:
given incompletely sampled
and noisy data $\bi{d}$ we want to reconstruct the true underlying field
$\bi{s}$ or true image $\bi{Rs}$
so that the reconstructed field is in some sense the closest to 
the real field. One way to approach this problem is to require the 
estimated field $\hat{\bi{s}}$ 
to be a linear function of the data, $\hat{\bi{s}}=
\bi{\Phi d}$, where $\bi{\Phi}$ is a $N \times M$ matrix. Then one can 
minimize
the variance of the residual 
\begin{equation}
\langle (\bi{s}-\hat{\bi{s}})
(\bi{s}-\hat{\bi{s}})^{\dag} \rangle
\label{res}
\end{equation}
with respect to $\bi{\Phi}$. This way  
we find the WF estimator of the underlying field,
\begin{equation}
\hat{\bi{s}}=\langle \bi{s}\bi{d}^{\dag}\rangle \langle \bi{dd}^{\dag}
\rangle^{-1}=\bi{SR}^{\dag} \bi{C}^{-1}\bi{d}.
\label{wf}
\end{equation}
The variance of residuals is given by
\begin{equation}
\langle \bi{rr}^{\dag} \rangle =\langle (\hat{\bi{s}}-\bi{s})
(\hat{\bi{s}}-\bi{s})^{\dag} \rangle=
\bi{S-S R}^{\dag}\bi{ C}^{-1} \bi{R S}.
\label{var}
\end{equation}
The reconstruction here is written in term of mode expansion 
coefficients. To make a 
real space map we transform them back to 
the real space, giving reconstructed image 
$\hat{\bi{d}}=\bi{R}\hat{\bi{s}}$.

For gaussian random fields Wiener filtering coincides with 
maximum probability  
estimator (\cite{zar}) and so it is optimal. 
To show this one can write the joint probability of signal $\bi{s}$
and noise $\bi{n}$ as a product of individual probabilities, under
the assumption that they are uncorrelated,
\begin{equation}
P(\bi{s},\bi{n})=P_s(\bi{s})P_n(\bi{n})=
(2\pi)^{-(M+N)/2} \det(\bi{S})^{-1/2} \det(\bi{N})^{-1/2} \exp(-{1 \over 2}
[\bi{s}^{\dag} \bi{S}^{-1}\bi{s}+\bi{n}^{\dag} \bi{N}^{-1}\bi{n}]).
\label{psn}
\end{equation}
For simplicity we have assumed that the noise from aliasing of
unestimated modes is included in the noise matrix $\bi{N}$. 
We will derive this term explicitly in equation (\ref{fsd}). 
The conditional 
probability for $\bi{s}$ given the data is $P(\bi{s} \vert \bi{d})
\propto P_s(\bi{s})P_n(\bi{d-Rs})$, which after completing the square
gives 
\begin{equation}
P(\bi{s} \vert \bi{d}) \propto \exp(-{1 \over 2}[\bi{s}-\hat{\bi{s}}]^{\dag}
(\bi{S}^{-1}+\bi{R}^{\dag} \bi{N}^{-1}\bi{R})[\bi{s}-\hat{\bi{s}}]),
\label{psd}
\end{equation}
where $\hat{\bi{s}}$ is given in equation (\ref{wf}). The covariance
matrix of residuals 
$\bi{S}^{-1}+\bi{R}^{\dag} \bi{N}^{-1}\bi{R}$ is identical to the one
in equation (\ref{var}), as can be obtained
by matrix manipulation using the 
matrix expansion $(\bi{I}+\bi{X})^{-1}=(\bi{I-X+XX-}...)$. The 
posterior probability for $\bi{s}$ 
is therefore maximized by WF solution
$\hat{\bi{s}}$.

In the above example we wanted to 
find the most probable field $\hat{\bi{s}}$ given the data $\bi{d}$. 
On the other hand, in previous section we wanted 
to maximize the probability  or likelihood
function of the data as a function of parameters $\Theta_l$
independent of the underlying field $\hat{\bi{s}}$. In this 
case one wants to marginalize over the parameters $\hat{\bi{s}}$ 
(\cite{rp}),
\begin{eqnarray}
P( \bi{d}) &=& \int P(\bi{s},\bi{d-Rs}) d^M\bi{s}  \nonumber \\
&=&(2\pi)^{-(M+N)/2} \det(\bi{S})^{-1/2} \det(\bi{N})^{-1/2} 
\exp\{-{1 \over 2}\bi{d}^{\dag}\bi{C}^{-1}\bi{d} \}
 \times \nonumber 
\\ &\int & \exp\{-{1 \over 2}[\bi{s}-\hat{\bi{s}}]^{\dag}
(\bi{S}^{-1}+\bi{R}^{\dag} \bi{N}^{-1}\bi{R})[\bi{s}-\hat{\bi{s}}]\}d^M\bi{s}
\nonumber \\
&=& (2\pi)^{-N/2} \det(\bi{C})^{-1/2} \exp(-{1 \over 2}
\bi{d}^{\dag} \bi{C}^{-1}\bi{d}).
\label{marg}
\end{eqnarray}
We see that to marginalize over a set of parameters $\bi{s}$ 
one has first to
find  their maximum probability value $\hat{\bi{s}}$, 
after which the integration 
over the parameters can be trivially performed.  
In this example the final result does not depend on the underlying field, 
and we could have directly written the probability distribution 
(or the likelihood function) for the data given the theory, as
we did in the previous section (the derivation here merely provides a 
more rigorous justification for the statement in the footnote 1). In 
other cases (e.g. when the data depend nonlinearly on the field) 
going through this intermediate step is necessary.

WF only uses information
on the mean and variance of statistical distribution, which 
completely characterize gaussian random fields and so is 
optimal for such applications. On the other hand, by using only
this information it may be less than optimal when applied 
to a strongly nongaussian field.  A typical example is image
reconstruction from incompletely sampled data in Fourier space, when
the image has significant point sources. In this case
nonlinear methods, such as maximum entropy method, are the methods
of choice, since they allow one to extract better the smooth component
underneath the ripples produced by point sources (\cite{nar}). But in the 
applications to LSS or CMB with small or no 
deviations from gaussianity mean and variance contains all the 
information on the statistical properties of the 
field and WF is the optimal method. It should be emphasized 
that even in nongaussian situations WF still minimizes the variance
as defined in equation (\ref{res})
among the class of linear estimators.
For example, for cluster 
reconstruction from weak lensing there exist several methods that
are linear in the data (\cite{sk}), 
but since WF explicitly 
minimizes the variance in equation (\ref{res}) it is in 
this sense optimal in this class of reconstruction 
methods (\cite{seljak97b}). However, typically 
in such applications minimizing 
the variance may not be the only way to define the best image
reconstruction and in general there is no optimal reconstruction
for all purposes.

To make a connection between WF and
the minimum variance estimator we rewrite equation (\ref{estimator}) as
\begin{equation}
(\bi{F}\hat{\bi{\Theta}})_l={1 \over 2}[(\bi{\Pi}_l\bi{R}^{\dag} 
\bi{C}^{-1}\bi{d})^{\dag}
(\bi{\Pi}_l\bi{R}^{\dag} \bi{C}^{-1}\bi{d})
-b_l].
\label{mv}
\end{equation}
By comparing equations (\ref{wf}) and (\ref{mv}) one finds
that the minimum variance power spectrum 
estimator can be simply expressed in terms of WF reconstructed field,
\begin{equation}
(\bi{F}\hat{\bi{\Theta}})_l=
{1 \over 2}[\bi{\Pi}_l\bi{S}^{-1}\hat{\bi{s}}^{\dag}
\hat{\bi{s}}\bi{S}^{-1}\bi{\Pi}_l-b_{l}],
\label{wfmv}
\end{equation}
where 
\begin{equation}
b_l=tr(\bi{\Pi}_l\bi{R}^{\dag} \bi{C}^{-1}\bi{(N+C_b})\bi{C}^{-1}\bi{R}\bi{\Pi}_l)
\label{wfbl}
\end{equation}
and 
\begin{equation}
F_{ll'}= {1 \over 2} tr(\bi{C^{-1}Q}_l\bi{C^{-1}Q}_{l'})={1 \over 2}
\sum_{m_l} \sum_{m_{l'}} |\bi{\Pi}_l\bi{R}^{\dag} \bi{C}^{-1}\bi{R} \bi{\Pi}_{l'}|_{m_lm_{l'
}}^2.
\label{wffll}
\end{equation}
These expressions provide additional insight to the minimum variance 
power spectrum estimator and its relation to the WF. To compute the
modes both methods
first weight the data by multiplying it with the inverse covariance 
matrix $\bi{s}=\bi{R}^{\dag}\bi{C}^{-1}\bi{d}$. This is not
suprising, since it is 
just a generalization of the usual inverse variance
weighting of the data, where now a given measurement can be downweighted
either because it has a large measurement error or because it is 
strongly correlated with other measurements (in which case it does not
provide additional information). Minimum variance
power spectrum estimator then simply averages over 
all the modes contributing to the $l$-th parameter in
the spectrum, while WF filters those that are below the noise.
To properly normalize the power spectrum estimates and to
provide their covariance matrix we also need to
subtract the noise bias $b_l$ and compute
the Fisher matrix $\bi{F}$.
This estimator therefore improves upon
the naive power spectrum estimation obtained by simple average 
$M_l^{-1}(\bi{\Pi}_l\hat{\bi{s}}^{\dag}
\hat{\bi{s}}\bi{\Pi}_l)$, which leads to biased results 
in the low signal to noise regimes (e.g. \cite{bunn}). 
It is also the best possible estimator, as discussed in previous 
section. Clearly it should be used whenever WF on the data is 
performed and one wants to obtain a power spectrum estimate as well. 
Since WF requires power spectrum as input in 
the absence of any external information one 
can use the actual  
power spectrum as measured from the data itself for WF 
reconstruction (or a smoothed version of it as discussed in section 2). 
This gives a self-consistency to the WF reconstruction
in the sense that it does not have to rely on any ad-hoc assumptions
when there is no external information.
Since WF is basically multiplying the modes with signal/(signal+noise)
one can 
see that the modes with no statistically significant excess 
of power are being filtered out and replaced with zero. Only the modes
where the power spectrum does show an excess of power above the noise 
will be kept in the reconstruction, thus providing the most conservative
reconstruction that is consistent with the data. For gaussian random 
fields this is in fact the optimal reconstruction. 

The most expensive operation in computing WF is inverting 
$\bi{C}$ and once this inversion is obtained it is straightforward
to compute $b_l$ and $\bi{F}$, so computing the power spectrum from 
WF estimators is not significantly more expensive than computing WF itself.
This inversion is O($N^3$) operations, if one uses
Cholesky decomposition \footnote[2]{Iterative procedures such as multigrid
methods coupled with Jacobi or Gauss-Seidel iteration can in principle
provide significant speedups for very large systems.
Such methods could be used for WF, but not for
power spectrum estimation, because to compute
$\bi{b}$ or $\bi{F}$ one has to solve $\bi{C}^{-1}\bi{R}$, meaning
the iteration has
to be repeated $M$ times and one is better off computing
$\bi{C}^{-1}$ explicitly. A possible solution is to compute 
$\bi{C}^{-1}\bi{d}$ using an iterative method and repeat 
this for a set of Monte Carlo realizations of the data and 
noise. By averaging over these one can compute 
$\bi{F}$ and $\bi{b}$ directly. } 
This is not computationally feasible for systems larger than $N \sim 10^4$
and so it is useful to reduce the size of the system when possible.
We discuss this in the next section.

\section{Signal eigenmode representation}
As mentioned in previous section the computational cost of performing 
both WF and minimum variance power spectrum estimation scales as $O(N^3)$.
Clearly reducing the size of the 
matrix is the first simplification one should attempt to try. 
One way to achieve this is to note
that the data are often
oversampled: in the case of COBE for example there are about 4000
pixels after the galactic cut, but only about 900 modes are required to
fit these data, as COBE has no information about the modes 
above $l \sim 30$. Similarly, if signal to noise for each measurement
is small (as is the case in weak lensing, with intrinsic ellipticity
of individual galaxies being much larger than the signal we are after),
then the number of data points will in general be much
larger than the number of modes that can be extracted. This suggests
to transform the data to the signal eigenmode basis first and perform
all the operations there. For computational reasons we will
do this by dividing the data first with the inverse of the noise matrix,
so that our data in the signal eigenmode basis are $\tilde{\bi{d}}=
\bi{R}^{\dag} \bi{N}^{-1} \bi{d}$ \footnote[3]{This is computationally
advantageous only if noise matrix is diagonal. If this is not the 
case then one can perform the transformation directly on the data,
$\tilde{\bi{d}}= \bi{R}^{\dag} \bi{d}$.
This leads to a similar set of expressions to those given
in the text and will not be explicitly presented here.}.
Using this vector as a new data set we can derive analogous 
expressions to equations (\ref{wf}), (\ref{wfbl}) and (\ref{wffll}),
\begin{equation}
\hat{\bi{s}}=[\bi{S}^{-1}+\bi{S}^{-1}(\bi{R}^{\dag} \bi{N}^{-1}\bi{R})^{-1}
\bi{R}^{\dag} \bi{N}^{-1}\bi{C_b}\bi{ N}^{-1}\bi{R}+
\bi{R}^{\dag} \bi{N}^{-1}\bi{R}]^{-1}
\tilde{\bi{d}} \equiv \bi{D}^{-1}
\tilde{\bi{d}},
\label{fsd}
\end{equation}
\begin{equation}
b_l=
tr[\bi{\Pi}_l\bi{S}^{-1}\bi{D}^{-1}\bi{R}^{\dag}(\bi{N}^{-1}\bi{C_b}\bi{N}^{-1}+\bi{N}^{-1})\bi{R}
\bi{D}^{-1}\bi{S}^{-1}\bi{\Pi}_l]
\end{equation}
and
\begin{equation}
F_{ll'}= 
\sum_{m_l} \sum_{m_{l'}} |\bi{\Pi}_l\bi{S}^{-1}\bi{D}^{-1}\bi{R}^{\dag} 
\bi{N}^{-1} \bi{R}\bi{\Pi}_{l'}|_{m_lm_{l'}}^2,
\end{equation}
whereas equation (\ref{wfmv}) remains unchanged.

The role of the correlation matrix $\bi{C}$ in real space has 
now been replaced by $\bi{D}$, which has dimension $M \times M$. 
If $M$ is significantly smaller than $N$ then a substantial saving in 
computational time can be achieved by having to invert a smaller matrix. 
This matrix has three contributions, signal, aliasing and noise terms.
By using this expression
we also need to compute $\bi{R}^{\dag} \bi{N}^{-1}\bi{R}$, which in 
principle requires $M^2N$ multiplications, although 
if one can take advantage of plane wave expansion 
and fast Fourier transforms (or their analogs for different basis 
functions)
then this can be significantly speeded up 
(see next section).
To compute the aliasing term this matrix needs to be inverted, which 
is $O(M^3)$, but to compute the complete aliasing term is in fact
O($MN^2+M^2N$), which becomes the most expensive operation. Fortunately,
this term can often be neglected, in which case computing $\bi{D}$
simplifies further (it also makes it symmetric and hence it can 
be inverted with Cholesky rather than LU decomposition). Matrix
$\bi{D}^{-1}$ also gives the variance of residuals as shown in 
equation (\ref{psd}) 
(where now the generalized noise term has been explicitly divided 
into aliasing and noise contributions). 
Same manipulations also reduce the computational cost to 
compute $\bi{b}$ and $\bi{F}$ as well.
All the expensive operations scale  
with $M$ only and there is no O($N^3$) left in the problem. 

The advantage of this transformation 
has been noted for WF applications (e.g. \cite{bunn}), but obviously
it is equally useful for minimum variance power spectrum estimation. 
The computational advantage of signal eigenmode representation
is particularly important for weak lensing or CMB polarization analysis,
if only one of the two independent modes is excited (as is the case
of weak lensing and often also for CMB polarization). Then by 
transforming to signal eigenmodes the size of the system is immediately
reduced by a factor of two (and hence the computational cost by a factor 
of eight), even if we use the same number of eigenmodes
as the number of data points (\cite{seljak97b}). 
A further advantage in using the expressions above is that
no pixelization of the data is necessary. This is because the response
matrix $\bi{R}$ can be evaluated at the exact positions of the measurements
and so the projection of the data to a given mode involves no approximations.
This would be particularly important if one wants 
to determine the power spectrum on small scales,
where pixelization can introduce significant errors in the method. 
The only potential disadvantage of working in this basis 
is that the matrix $\bi{D}$ can be 
singular, even though it is regularized with the addition of $\bi{S}^{-1}$.
This happens for example when the data are sparse or there are 
large unobserved regions in the survey. This is not so much 
a problem for WF, since one can
use Singular Value Decomposition (SVD, \cite{Press92}) to 
decompose the matrix and zero all the small eigenvalues, which would
destabilize the inversion. One cannot however use this to estimate
the power spectrum, because it leads to biased results. 

\section{Plane wave expansion}
We now apply the formulae derived in previous sections to the 
case where the eigenmodes are plane waves. This is the case of 
significant importance, as it can be applied to the analysis of 
galaxy survey data (in 1-d, 2-d and 3-d), 
weak lensing, Ly-$\alpha$ forest and small scale CMB anisotropies. 
We will assume that the 
noise is uncorrelated between the measurements, as is usually the case
here. There are two simplifications
with respect to the general case: first is if noise is diagonal, 
which allows one to compute 
$\bi{R}^{\dag} \bi{N}^{-1} \bi{R}$ with O($MN$) operations instead 
of O($M^2N$). Second is the use of FFT, 
which further reduces the 
operational count of this operation to  O($N \ln N$). 
We also discuss 
an approximation scheme that can be developed for the solution, 
allowing very large systems to be solved. Although analogous approximation
can also be developed for the general case, the two advantages of plane
wave expansion make it particularly suitable to apply it in this case.

Given the data 
$\bi{d}$ we first interpolate them to a grid with
$N_g$ points in each of $N_d$ dimensions.
The size of the grid $D$ in each dimension has to be such that it covers
all the data available \footnote[4]{Since we are sampling the field 
in a finite volume one can only determine the power spectrum convolved
with the window of the data and one cannot resolve it better than 
$\Delta k \sim 2\pi/D$. If the power spectrum is smooth over $\Delta k$
then this 
additional aliasing from nearby modes can be ignored.}, while the grid spacing should be at least 
one half of 
the smallest scale we hope to resolve with the data. The reason for 
this choice will become apparent later. We will assume here for 
simplicity that the interpolation is simply assignment
to the nearest grid point. 
This is not essential because as we will show all 
the operations scale as O($N \ln N$), where $N=N_g^{N_d}$ 
is the number of pixels and $N_d$ is the dimension of the survey,
so one can easily increase the 
number of pixels with only (almost) linear
increase in computational time. 
If there are more than one 
data point assigned to a given pixel then we average over all the 
points by inverse noise weighting. If all the data have equal noise 
variance $\sigma^2$
then this amounts to simple averaging of the data, $d_{\bi{r}}=
N_{\bi{r}}^{-1}
\sum_{k=1}^{N_{\bi{r}}}d_k$, 
where the sum is over all $N_{\bi{r}}$ data points that contribute 
to $\bi{r}$-th grid point. 
The variance at that point is given 
by $N_{\bi{r}}^{-1}\sigma^2$, where $\sigma^2$ is the noise 
variance for each 
measurement. 
The data are now represented with the values $d_{\bi{r}}$ and the 
corresponding noise variance $\sigma^2_{\bi{r}}=N_{\bi{r}}^{-1}\sigma^2$. 
We can define two new vectors, 
inverse noise weighted data
$h_{\bi{r}}=d_{\bi{r}}/\sigma_{\bi{r}}^2$ and inverse noise 
$\sigma^{-2}_{\bi{r}}$. 
Note that if a certain grid point $\bi{r}$ has no galaxies contributing
to it then the corresponding values $d_{\bi{r}}$, $h_{\bi{r}}$
and $\sigma^{-2}_{\bi{r}}$ 
are set to 0 so all the points on the grid have a well defined value.

The response matrix $\bi{R}$ for the plane wave expansion 
is $R_{\bi{k,r}}=e^{i\bi{k}\cdot\bi{r}}$,
where $\bi{k}$ and $\bi{r}$, the Fourier mode and real space
position in the box respectively, can be parametrized with 
$N_d$-dimensional indeces $\bi{l}$ and $\bi{x}$. 
The Fourier mode $\bi{k}$ 
is obtained from its index as $\bi{k}=2\pi \bi{l}/D$, while the real 
space position is related to the index as $\bi{r}=\bi{x}D/N_g^{N_d}$, 
where $D$ is the size of the grid. While each component of $\bi{x}$ runs 
between 0 and $N_g-1$, components of $\bi{l}$ run between $-N_g/2$ and 
$N_g/2$, with the two extreme values being equal 
(corresponding to the Nyquist frequency). 
If the data are non-scalar quantities then 
the above expression has to be generalized by
adding a function 
$\chi(\bi{k})$ to the response matrix $\bi{R}$. For example,
if the data are two shear components in the case of weak lensing then 
$\chi(\bi{k})=(\cos{2\phi_k},\sin{2\phi_k})$ for the two shear 
components, where $\phi_k$ is the
phase of the Fourier mode $\bi{k}$. We will ignore this factor
in the following, as it can easily be added to the final expressions.
Projecting the data to Fourier 
space as in the previous section gives
\begin{equation}
\bi{R}^{\dag}\bi{N}^{-1}\bi{d}=\sum_i e^{i\bi{k}\cdot \bi{r}_i}{
d(\bi{r}_i) \over 
\sigma^2(\bi{r}_i)}\approx \sum_{\bi{r}}
e^{i\bi{k}\cdot\bi{r}}{ d_{\bi{r}} \over \sigma_{\bi{r}}^2}=
\sum_{\bi{r}}
e^{i\bi{k}\cdot\bi{r}}h_{\bi{r}}=\tilde{h}_{\bi{k}},
\end{equation}
where we denoted with $\tilde{h}_{\bi{k}}$
the Fourier transform of $h_{\bi{r}}$, which can be computed using FFT. 
Note that
the only approximation in the above expression was to replace the real 
positions of measured data 
with their interpolated position in the grid and as mentioned above 
this can 
always be made more accurate by increasing the dimension of the grid.
Since computing the Fourier transform with FFT is O($N\ln N$)
operations the computational time is small even for very
large ($\sim 10^8$) matrices. Of course, one can always perform the 
direct summation over all the individual data points, in which 
case the operation count will be O($MN$) with no approximation
because of pixelization and may be more advantageous when $M$ is
small.

Next we need to compute the matrix 
$\bi{D}$ (equation \ref{fsd}).
$\bi{S}^{-1}$ is diagonal in Fourier space, 
while noise transformation can be 
written as
\begin{equation}
(\bi{R}^{\dag} \bi{N}^{-1}\bi{R})_{\bi{k_1,k_2}}=
\sum_i {e^{i\bi{k_1}\cdot \bi{r_i}} e^{-i\bi{k_2}\cdot \bi{r_i}}
\over \sigma^2(\bi{r}_i)} \approx \sum_{\bi{r}} e^{i(\bi{k}_1-\bi{k}_2)
\cdot\bi{r}}\sigma_{\bi{k}}^{-2}=(\tilde{\sigma}^{-2})_{\bi{k}_1-\bi{k}_2},
\end{equation}
which is the Fourier transform of $\sigma^{-2}_{\bi{r}}$ evaluated
at $\bi{k}_1-\bi{k}_2$. This can again be obtained from FFT of the 
same grid as before, provided the initial grid was oversampled by a 
factor of 2 in each dimension, so that we are only interested in the
modes that are less than one half of Nyquist frequency of original grid. 
This part of calculation is therefore also O($N\ln N$) as advocated
in previous section, instead of being
O($M^2N$) (without pixelization to a grid and 
FFT it is O($MN$)). Finally, when required we also need to compute the
aliasing term in equation (\ref{fsd}). If only the small scale aliasing
is important it can be obtained by computing first
\begin{equation}
(\bi{R}^{\dag} \bi{N}^{-1}\bi{R}_b)_{\bi{k_1,k_2}}=
(\tilde{\sigma}^{-2})_{\bi{k}_1-\bi{k}_2},
\end{equation}
where $\bi{k}_2$ now corresponds to one of the modes we do not estimate.
We can then subsequently perform the matrix multiplications in 
equation (\ref{fsd}) to obtain the aliasing term. Of course, this requires 
that the dimension of the grid is sufficiently large to encompass all
of the modes that alias power into the modes we wish to estimate, so
one should always check to see if the results change by changing the 
dimension of the grid.
When this becomes too large one is better off to compute the 
correlation matrix due to unwanted modes 
in real space analytically and then project it 
to Fourier space. Because the correlation matrix is not diagonal one
cannot perform this with a single FFT but rather with $N$ of these
and operation count becomes $O(N^2\ln N)$. Once this is done then
everything has been 
projected to Fourier space and all the subsequent 
operations scale with $M$ 
instead of $N$. In particular, inverting $\bi{D}$ 
and computing $\bi{b}$ and $\bi{F}$ 
are all O($M^3$) operations.

For very large $M$ performing these matrix manipulations
become computationally too
expensive, so it is worth exploring approximations to the above 
expressions, which would allow one to compute them more rapidly. 
When the data are Poisson sampled one can simplify the analysis 
by noting that to weight the data one can 
use the mean number of measurements instead of the actual value, i.e. 
$\sigma^2_{\bi{r}}=N_{\bi{r}}^{-1}\sigma^2$ is being replaced with
$\sigma^2_{\bi{r}}=\bar{N}_{\bi{r}}^{-1}\sigma^2$ and 
$d_{\bi{r}}=
N_{\bi{r}}^{-1}
\sum_{k=1}^{N_{\bi{r}}}d_k$ with $d_{\bi{r}}=
\bar{N}_{\bi{r}}^{-1}
\sum_{k=1}^{N_{\bi{r}}}d_k$, where $\bar{N}_{\bi{r}}$
is the mean density of measurements. We continue to keep the subscript
$\bi{r}$ attached to this quantity, since it may vary across the survey
(for example because of varying selection function as a function of 
distance in galaxy surveys).
If we assume a 
white noise power spectrum, $\Theta_l={\rm const}$ for all $l$, then
one can solve the system exactly without 
any inversion. The reason is that the covariance matrix $\bi{C}$ becomes 
diagonal in real space, $C_{\bi{r}\bi{r}'}=(\sigma^2+\sigma_s^2
)\delta_{\bi{r}\bi{r}'}$, where $\sigma_s^2=\Theta_l\bar{n}(\bi{r})$ 
is the theoretical variance, with $\bar{n}(\bi{r})$ being the mean 
density of measurements at location $\bi{r}$. 
This matrix 
can be inverted trivially in real space and $h_{\bi{r}}=(\bi{C}^{-1}\bi{d})_{\bi{r}}=
d_{\bi{r}}/(\sigma^2+\sigma_s^2)$ 
is a simple inverse weighting of the data, where the weight 
consists now of both noise and signal variance. 
If the survey is compact then one can locally approximate
the power spectrum as white noise and make $\bi{C}$ diagonal,
but vary theoretical variance $\sigma_s^2$ 
using the actual power spectrum $\Theta_l$. This 
corresponds to the Feldman, Kaiser and Peacock (1994) weighting, which is 
exact for the power spectrum estimation in the so-called
classical limit, where the wavelength of interest is much shorter than 
the size of the survey (\cite{hama,tegb}).
WF estimator becomes
\begin{equation}
\hat{s}_{\bi{k}}=\Theta_l \tilde{h}_{\bi{k}},
\end{equation}
with the variance of residuals
\begin{equation}
\langle r_{\bi{k}}r^*_{\bi{k'}} \rangle =
\Theta_l[\delta_{\bi{k}\bi{k'}}-\Theta_l\tilde{\bi{C}}^{-1}(\bi{k}-\bi{k'})].
\end{equation}
The bias and Fisher matrix are given by
\begin{eqnarray}
b_l=M_l \sum_{\bi{i}} { \sigma_{\bi{r}_i}^2\over (\sigma^2+\sigma_s^2)^2}
\nonumber \\
F_{ll'}=\sum_{\bi{k}}\sum_{\bi{k'}} 
|\tilde{\bi{C}}^{-1}(\bi{k-k'})|^2.
\end{eqnarray}
Here $\bi{k}$, $\bi{k'}$ are the wavevectors corresponding 
to parameters $\Theta_l$ and $\Theta_{l'}$, respectively, and 
$\tilde{\bi{C}}^{-1}$ is a Fourier transform of $(\sigma^2+
\sigma_s^2)^{-1}$.
We see that
a complete solution to the problem can be obtained 
without performing any matrix 
inversion and all the operations are O($N \ln N$), a huge
advantage over the general case. Because of this   
we implicitly assumed all the relevant modes are being estimated 
and ignored the aliasing term. If $\bar{n}$ is not varying with $\bi{r}$
then the weighting of the data is uniform and this gives a simple 
criterion for signal to noise of the modes given by 
$\bar{n}\Theta_l/\sigma$. 
One cannot take advantage of FFT to transform $\bi{C}^{-1}$ directly 
in this case because of $k$ 
dependence in real space. Instead one can do a couple of FFT varying 
$\sigma_s^2$ with $k$ (for example one FFT for each $l$ using $\Theta_l$
to compute $\sigma_s^2$) 
and then combining them into a single matrix so that 
for a given pair $\bi{k}$, $\bi{k}'$ in $\tilde{\bi{C}}^{-1}$
one chooses $\Theta_l$ that corresponds to $\bi{k}$ (\cite{hama}). 
The computational cost is still approximately O($N \ln N$) and
the error is negligible provided the Fisher matrix is nearly 
diagonal, which is the assumption that goes into the classical limit.
This procedure has been applied to the
power spectrum estimation in the case of galaxy surveys (\cite{fkp}), but 
one can apply it to other data sets and to the case of WF reconstruction 
as well, whenever the survey geometry is compact and the data sampling 
relatively uniform. 
One can further improve the 
reconstruction by performing the full analysis for the largest modes,
where the classical approximation is the least reliable and use
the classical approximation on smaller scales.  

\section{Discussion}
In this paper we explore the close 
connection between optimal methods for image reconstruction 
and minimum variance
power spectrum estimator. We show that for the case of gaussian random 
fields both can be obtained within the same formalism, providing a unifying
link between the two. This is because the first step, inverse variance 
weighting of the data, is common to both methods as one would expect: 
if a given measurement has a large noise or is strongly correlated with
the other measurements then it does not add new information and it should
be downweighted by the inverse of covariance matrix. Once this operation
is performed then the power spectrum estimator simply averages over all
the modes contributing to a given spectral bin, while WF filters the 
mode according to the signal/noise ratio. 
If the requirement
of gaussianity is dropped then the two methods remain useful approximations.
WF cannot claim to be optimal 
for all applications, although by definition it still minimizes the 
variance among all the linear functions of the data and therefore 
continues to be useful in many instances.  Similarly, while
there exists a minimum variance power spectrum estimator (\cite{hama}),
its expression is generally too complicated to compute and the gaussian 
approximation remains useful. One has to be careful in this case to  
provide proper error estimates, which in general should be done 
either using second order perturbation theory or using 
N-body Monte Carlo simulations. Alternatively, one can use 
a quick (and dirty) Monte Carlo estimate of error bars by bootstraping 
the WF estimators $\bi{s_l}$ contributing to a given power spectrum 
bin $l$. This means drawing a set of WF
estimates from the original estimated set with a replacement, so that 
some of the values will be drawn several times, while others none.
One then repeats the power spectrum 
estimation procedure for each drawn set and the 
scatter in the estimators gives an estimate of the error.
To the extent that the scatter in the power spectrum
estimates is indicative of the error on their average this will give  
reasonable results.
However, for bootstrap method to be formally applicable one needs the
data to be independent and identically distributed and this is 
not necessarily the case because of noise, sparse sampling and nonlinear 
evolution. Moreover, bootstraping will underestimate the error where there
are not enough independent realizations of a given mode amplitude. 
This is mainly a problem on large scales, where 
a gaussian error estimate often suffices. 
Clearly this procedure needs to be carefully tested with N-body simulations
before it can be recommended, but it should at least 
superseed the error estimates based on the variance in subsamples of
the data (e.g. \cite{lcrs}). 

There is another advantage in relating the two procedures
that we attempted to stress in the paper. It is the 
possibility of translating various computational schemes 
devised for one method to the other. As an example,
WF in the signal eigenmode space  
is often faster than in the space of real data 
and while this has been explored in the 
past for WF analysis (\cite{bunn}) it can be applied in the same way
also to minimum variance power spectrum estimator.  
Conversely, approximations
such as the classical limit (\cite{fkp}) used for power spectrum
analysis can be applied to WF as well. In general there is a 
distinctive advantage of having a unified approach to the two 
problems, since finding a fast solution to one problem will 
also lead to a fast solution to the other problem. 
Finally, being able to compute the power spectrum estimate also 
answers the question what power spectrum to use in WF in the absence 
of any prior information:
one simply uses the minimum variance 
power spectrum measured from the data itself
or one that is consistent with it
(e.g. a smoothed version of this power 
spectrum). We hope this will provide a unified
frame to many seemingly unrelated statistical approaches and
will add a further incentive to those analyzing the data
to use these powerful statistical methods on existing and 
future surveys. 

\acknowledgements
I would like to thank Lam Hui and the participants 
at the Aspen workshop on Large Scale Structure for useful conversations.
I would also like to acknowledge the hospitality of the Aspen Center 
for Physics, where part of this work was completed.

\appendix
\section{Appendix: External constraints}

Minimum variance power spectrum 
estimator and WF in the form we presented in the text have to be modified
when the data are contaminated or do not contain information for some
modes, when additional constraints are placed on the data or 
when there are additional external parameters not included in the 
model. For example, in CMB data monopole and 
dipole are contaminated by mean temperature and 
by our motion with respect to the CMB frame, respectively.
Because of incomplete coverage these modes contaminate higher modes
as well and have to be removed from the data set.
Similarly, in LSS galaxy surveys the mean density is unknown and 
obtained from the measured data itself, so $\bi{k}=0$ mode
mixes and contaminates the higher $\bi{k}$
modes as well if the modes are coupled. Other constraints which are 
not in the form of one of the modes can be placed on the data as well. 
For example, one may first want to remove a quadratic or 
cubic trend from the 
Ly-$\alpha$ forest to account for the variable QSO continuum,
before reconstructing the density field or its power spectrum. Although in 
some instances we want to remove these components from the data 
without knowing their actual values, in other cases we may be 
interested also in their best reconstructed values. We also need to
specify what do we mean by removing a certain component. One possibility
is to project the data to a subspace orthogonal to these modes,
so that all the unreliable information in these modes will be 
destroyed (\cite{teg}). 
From a Bayesian point of view one wants instead to marginalize
over the modes we are not interested in 
by integrating over their probability distribution.
Although seemingly different all these cases lead to the same solution,
which can easily be incorporated into the formalism in the text. 
To show this let us first discuss the case of simultaneous 
reconstruction and external parameter determination. Following 
Rybicki \& Press (1992) we can write the data vector
 in the presence of external parameters 
as $\bi{d}=\bi{Rs}+\bi{Lq}+\bi{n}$ (again
we have implicity put the aliasing modes into the noise term), 
where $\bi{L}$ is the $N \times M_q$ matrix with known coefficients 
and $M_q$ is the number of external parameters. 
For example, when determining the mean $\bi{L}=(1,1,...,1)$.
If we assume a uniform 
prior for $\bi{q}$ ($P_q(\bi{q}) \propto \rm{ const}$) then 
\begin{equation}
P(\bi{q},\bi{s},\vert \bi{d}) \propto P_s(\bi{s})P_n[\bi{d}-
(\bi{Rs}+\bi{Lq})]
\end{equation}
We can now simulteneously extremize the posterior probability for 
$\bi{q}$ and $\bi{s}$. This gives 
\begin{eqnarray}
\hat{\bi{q}}&=& (\bi{L}^{\dag} \bi{C}^{-1} \bi{L})^{-1}
\bi{L}^{\dag} \bi{C}^{-1} \bi{d}, \nonumber \\
\hat{\bi{s}}&=& \bi{SR}^{\dag} \bi{C}^{-1}
(\bi{d}-\bi{L}\hat{\bi{q}}).
\end{eqnarray}
If we wish to know the probability distribution of the data independent
of external parameters and the underlying field we marginalize over
these parameters by integrating $P_s(\bi{s})P_n[\bi{d}-
(\bi{Rs}+\bi{Lq})]$ over $d^{M_q}\bi{q} d^M \bi{s}$ 
as we did in equation (\ref{marg}).
This gives (\cite{rp,zar})
\begin{equation}
P(\bi{d}) \propto
\exp\left(-{1 \over 2}
\bi{d}^{\dag} [\bi{C}^{-1}-\bi{C}^{-1}\bi{L}(\bi{L}^{\dag} 
\bi{C}^{-1} \bi{L})^{-1}\bi{L}^{\dag} \bi{C}^{-1}]\bi{d}\right).
\end{equation}
Although we presented here the Bayesian derivation
the maximum likelihood method (in the frequentist sense) leads to the 
exactly same likelihood function for the data  
(\cite{rp}). 
To compute the power spectrum in the 
presence of external parameters we therefore 
need to replace $\bi{C}^{-1}$ in 
previous sections with $\bi{C}^{-1}-\bi{C}^{-1}\bi{L}(\bi{L}^{\dag} 
\bi{C}^{-1} \bi{L})^{-1}\bi{L}^{\dag} \bi{C}^{-1}$ and then proceed as
before. Note that 
\begin{equation}
[\bi{C}^{-1}-\bi{C}^{-1}\bi{L}(\bi{L}^{\dag}
\bi{C}^{-1} \bi{L})^{-1}\bi{L}^{\dag} \bi{C}^{-1}]\bi{d}=
\bi{C}^{-1}(\bi{d}-\bi{L}\hat{\bi{q}})\equiv
\bi{C}^{-1}\bi{\Pi} \bi{d},
\end{equation}
which is simply removing the best estimated external parameters 
$\hat{\bi{q}}$ 
contribution from the data. We defined the projection operator $\bi{\Pi}$
which has the property that $\bi{\Pi L}=0$ and so 
we see that this is 
also equivalent to projecting the data to the subspace orthogonal
to the unwanted modes or external parameters (\cite{tegb}).
Subtracting the best estimated external parameter, 
orthogonalizing the data and marginalizing the data all lead to 
the same result, as advertised above. 

Using the Woodbury formula one can further simplify the new 
correlation matrix (\cite{rp})
\begin{equation}
\bi{C}^{-1}-\bi{C}^{-1}\bi{L}(\bi{L}^{\dag}
\bi{C}^{-1} \bi{L})^{-1}\bi{L}^{\dag} \bi{C}^{-1}=\lim_{\sigma^2 \rightarrow
\infty} (\bi{C}+\sigma^2 \bi{LL}^{\dag})^{-1},
\end{equation}
so instead of computing the expression on the left hand side above one
can add to $\bi{C}$ 
a term proportional to $\bi{LL}^{\dag}$ with a large variance (see also 
\cite{kbj}).
This is not surprising, since we assumed a uniform prior for $\bi{q}$ 
above which is equivalent to having a large variance in the 
correlation matrix for this parameter.
This is often computationally the simplest 
approach, specially when we are not really interested in the external
parameters themselves. The discussion here is another example
of the close connection
between WF and minimum variance power spectrum estimator. Both 
give the same result when dealing with external parameters or 
other linear constraints and again
a single solution solves both problems. 

\end{document}